\documentclass[aps,preprint,amssymb,amsmath,nofootinbib,eqsecnum]{revtex4}
 \begin{document}
 \title{Noncommutative Bayesian Statistical Inference from a wedge of a Bifurcate Killing Horizon}
 \author{Gavriel Segre}
 \homepage{http://www.gavrielsegre.com}
 \email{info@gavrielsegre.com}
 \date{25-4-2004}
 \begin{abstract}
  Expanding the remark5.2.7 of \cite{Segre-02} the noncommutative
  bayesian statistical inference from one wedge of a bifurcate
  Killing horizon is analyzed looking at its inter-relation with
  the Unruh effect
 \end{abstract}
 \maketitle
 \newpage
 \tableofcontents
 \newpage
 \section{Notation}
 \begin{tabular}{|c|c|}
   $ x \; = \, y $ & x is equal to y \\
   $ x \; := \: y $ & x is defined as y \\
   $ card(S) $ & cardinality of S \\
   $ f_{\star} $ & differential map of f \\
   $ f^{\star} $ & pull-back of f \\
   $ {\mathcal{L}}_{X} $ & Lie derivative w.r.t. X \\
   $ Is[( M \, ,\, g_{a b })] $ & isometry group of the space-time  $ ( M \, ,\, g_{a b
   }) $ \\
   $ \Gamma ( T^{(r,s)} M)  $ & sections of the (r,s)-tensor bundle
   over M \\
   $ I^{-}(S) $ & chronological past of the space-time's region  S \\
   $ I^{+}(S) $ & chronological future of the space-time's region S \\
   $ J^{-}(S) $ & causal past of the space-time's region S \\
   $ J^{+}(S) $ & causal future of the space-time's region S \\
   $ D^{-}(S) $ & past domain of dependence of the space-time's region S \\
   $ D^{+}(S) $ & future domain of dependence of the space-time's region S \\
   $ D(S) $ & domain of dependence of the space-time's region S \\
   $ S(A) $ & space of the states over a $ W^{\star}$-algebra
    \\
    $ \tau_{unbiased} $  & unbiased state  \\
   $ A^{W}_{( M \, ,\, g_{a b })} $ & Weyl algebra of the
   space-time $ ( M \, ,\, g_{a b } ) $ \\
   $ S_{H} ( A^{W}_{( M \, ,\, g_{a b })} )$  & Hadamard states over $  A^{W}_{( M \, ,\, g_{a b })}
   $ \\
   $ \sigma^{\omega}_{t}$ & modular group of the state $ \omega $ \\
   $ AUT(A) $ & automorphisms of the $ W^{\star} $-algebra A \\
  $ INN(A) $ & inner automorphisms of the $ W^{\star} $-algebra A \\
  $ OUT(A) $ & outer automorphisms of the $ W^{\star} $-algebra A \\
   $ GR-AUT( G \, , \, A) $ & automorphisms' groups of A representing
  G \\
  $ GR-INN( G \, , \, A) $ & inner automorphisms' groups of A representing
  G \\
  $ GR-OUT( G \, , \, A) $ & outer automorphisms' groups of A representing
  G \\
  $ \circledS $ & semi-direct product of groups \\ \hline
 \end{tabular}
 \newpage
 \section{Noncommutative Bayesian Statistical Inference}
 Noncommutative Bayesian Statistical Inference, as introduced by Miklos Redei in the $ 8^{th} $ chapter \textit{"Quantum conditional and quantum
conditional probability"} of \cite{Redei-98}, is based on the
following analysis.

Given a classical probability space $ ( X \, , \, \sigma \, , \,
\mu ) $ let us suppose to be a statistician having access only to
the partial information concerning the probability of an event $
B \in \sigma $ and whose goal is to estimate the unknown
probability  $ \mu(A) $ of an arbitrary event $  A \in \sigma $.

The Bayesian recipe prescribes that, before using even the
partial information he has, the more natural estimation of $
\mu(A) $ is the one introducing no bias, i.e.:
\begin{equation}
  \mu_{A \; PRIORI} (A) \; := \; P_{unbiased} (A)
\end{equation}
i.e. the uniform distribution over $ (X , \sigma )$ whether $
card( X ) < \aleph_{0} $ or the Lebesgue measure over $ (X ,
\sigma )$ whether $ card( X ) = \aleph_{1} $.

Let us observe that in the case $ card (X) \; = \; \aleph_{0} $
the unbiased probability measure over $ (X , \sigma )$ doesn't
exist so that the Bayesian strategy of statistical inference is
not defined in that case.

The acquisition of the partial information he can access results,
according to the Bayesian recipe, in the following ansatz:
\begin{equation} \label{eq:commutative Bayesian recipe}
  \mu_{A \; PRIORI} (A) \; := \; P_{unbiased} (A) \; \rightarrow \; \mu_{A \; POSTERIORI}
  (A) \, := \, \frac{\mu_{A \; PRIORI}( A \bigcap B )}{ \mu(B) }
\end{equation}
Let us now recall the Basic Theorem of Noncommutative Probability
stating that the category having as objects the classical
probability spaces and as morphisms their automorphisms is
equivalent to the category having as objects the algebraic
commutative probability spaces and as morphisms their
automorphisms.

Such a theorem naturally leads to a noncommutative generalization
of the Bayesian recipe consisting in:
\begin{enumerate}
  \item the recasting of eq:\ref{eq:commutative Bayesian recipe}
  in the language of algebraic probability spaces
  \item the generalization to noncommutative probability spaces
\end{enumerate}

Given the algebraic commutative probability space  $ ( A \, , \,
\omega) $ with:
\begin{equation}
  A \; := \; L^{\infty} (X \, , \, \sigma \, , \, \mu)
\end{equation}
\begin{equation}
  \omega (a) \; := \; \int_{X} d \mu \, a \; \; a \in A
\end{equation}
let us suppose that the statistician has access only to the
information concerning a sub-$\sigma$-algebra $
\sigma_{accessible} $ of $ \sigma $.

Introduced the $ W^{\star}-algebra $:
\begin{equation}
  A_{accessible} \; := \;  L^{\infty} (X \, , \, \sigma_{accessible} \, , \, \mu_{accessible})
\end{equation}
where:
\begin{equation}
 \mu_{accessible} \; := \; \mu|_{\sigma_{accessible}}
\end{equation}
and the associated state $ \omega_{accessible} \in S(
A_{accessible} ) $:
\begin{equation}
  \omega_{accessible} (a) \; := \; \int_{X} d \mu_{accessible} \, a \; \; a \in A_{accessible}
\end{equation}
we can express the Bayesian recipe in the following way:
\begin{enumerate}
  \item before using even the partial information that is
  accessible to him, the better \textbf{a-priori estimation} of  $ \omega $
  the statistician can perform consists in introducing no bias, assuming that:
\begin{equation*}
  \omega_{A \; PRIORI} \; := \; \tau_{unbiased}
\end{equation*}
where $ \tau_{unbiased} $ is the unbiased state  over A :
\begin{equation}
  \tau_{unbiased} (a) \; := \; \int_{X}  d P_{unbiased} \, a \; \; a
  \in A
\end{equation}

  \item the adoption of the available information may be encoded
  in the passage from the \textbf{a-priori estimation} to the \textbf{a-posteriori
  estimation}  of $ \omega $ specified by the \textbf{ Bayes
  rule}:
\begin{equation}
   \omega_{A \; PRIORI} ( \cdot ) \, = \, \tau_{unbiased} ( \cdot ) \; \rightarrow \\
   \omega_{A \;
   POSTERIORI} ( \cdot ) \; := \;  \omega_{accessible} ( E_{A \; PRIORI} \cdot )
\end{equation}
where $ E_{A \; PRIORI} \, : \, A \mapsto A_{accessible} $ is the
\textbf{conditional expectation w.r.t. $ A_{accessible} $
$\omega_{A \; PRIORI}$-invariant}
\end{enumerate}

The Basic Theorem of Noncommutative Probability allows to
 generalize immediately such a recipe to the noncommutative case
in which a statistician has access only to the sub-noncommutative
probability space $ ( A_{accessible} \, , \, \omega_{accessible}
)$ of a larger noncommutative probability space $ ( A \, , \,
\omega ) $:
\begin{equation}
 \omega_{accessible} \; := \omega|_{A_{accessible}}
\end{equation}
resulting in the following \textbf{noncommutative bayesian
recipe}:
\begin{enumerate}
  \item before using even the partial information that is
  accessible to him, the better \textbf{a-priori estimation} of  $ \omega $
  consists in introducing no bias, assuming that:
\begin{equation*}
  \omega_{A \; PRIORI} \; := \; \tau_{unbiased}
\end{equation*}
where $ \tau_{unbiased} $ is the noncommutative unbiased
probability distribution over $ A $, namely the tracial state on
it.
  \item the adoption of the available information may be encoded
  in the passage from the \textbf{a-priori estimation} to the \textbf{a-posteriori
  estimation}  of $ \omega $ specified by the \textbf{noncommutative Bayes
  rule}:
\begin{equation}
   \omega_{A \; PRIORI} ( \cdot ) \, = \, \tau_{unbiased} ( \cdot ) \; \rightarrow \\
   \omega_{A \;
   POSTERIORI} ( \cdot ) \; := \;  \omega_{accessible} ( E_{A \; PRIORI} \cdot )
\end{equation}
where $ E_{A \; PRIORI} \, : \, A \mapsto A_{accessible} $ is the
\textbf{conditional expectation w.r.t. $ A_{accessible} $
$\omega_{A \; PRIORI}$-invariant}
\end{enumerate}
Let us now observe that to the feasibility condition for such a
statistical inference already present in the commutative case and
requiring the existence of the unbiased probability distribution,
another constraint has to be added in the noncommutative case:
according to Takesaki Theorem $ E_{A \; PRIORI} $ exists if and
only if the following  \textbf{modular constraint} is satisfied:
\begin{equation}
  \sigma^{\omega_{A \; PRIORI}}_{t} (a) \; \in \; A_{accessible}
  \; \; \forall a \in A_{accessible} \, , \, \forall t \in {\mathbb{R}}
\end{equation}
where $ \sigma^{\omega_{A \; PRIORI}}_{t} $ denotes the modular
group of $ \omega_{A \; PRIORI } $.

Let us observe, with this regard, that since the modular
constraint is not satisfied for any sub-$W^{\star}$-algebra, the
philosophical subjectivistic viewpoint consistent in the
commutative case, cannot be generalized to the noncommutative
case.

What may be generalized to the noncommutative case is the recipe
for statistical inference but not \emph{"one point of view in its
entirety"} \footnote{citing de Finetti's dedication of his
\emph{"Theory of Probability}" to a Segre I am not, unfortunately,
a descendant of}.

\newpage
 \section{Noncommutative Statistical inference from a wedge of a bifurcate Killing horizon}
Given a space-time, i.e. a 4-dimensional lorentzian manifold $ (
M \, , \, g_{a b } ) $ \footnote{I will follow Penrose abstract
index notation as explained, e.g., in \cite{Wald-84} }
\cite{Wald-84}, \cite{Wald-94}, \cite{Kobayaski-Nomizu-96}, let
us suppose that it admits a bifurcate Killing horizon, i.e. a
bidimensional spacelike surface S such that there exist a Killing
vector field $ X^{a} $ vanishing on it:
\begin{equation}
  {\mathcal{L}}_{X^{a}} g_{a b} \; = \; 0
\end{equation}
\begin{equation}
  X^{a}(p) \; = \; 0 \; \; \forall p \in S
\end{equation}

Let us suppose, furthermore, that S is a Cauchy surface of $ ( M
\, , \, g_{a b } )$ \footnote{We have implicitly assumed that $ (
M \, , \, g_{a b } )$ is globally-hyperbolic and, hence, admits
Cauchy surfaces. While in Classical General Relativity the status
of the Strong Cosmic Censorship Conjecture stating that any
"physical" space-time is globally hyperbolic is dubious, it is
strongly dubious whether a Quantum Field Theory on a non
globally-hyperbolic space-time may be consistently formalized},
i.e. that its domain of dependence is the whole M:
\begin{equation}
  D(S) \; = \; M
\end{equation}

Denoted by $ h_{A} $ and  $ h_{B} $ the two  null surfaces
generated by the null geodesics orthogonal to S, M may be
expressed as the union of four disjoint wedges:
\begin{align} \label{eq:splitting in four wedge}
   M \; & = \; \cup_{i=1}^{4} W_{i} \\
   W_{1} \;  &  := \; I^{-} ( h_{A} ) \, \cap \, I^{+} ( h_{B} ) \\
   W_{2} \;  &  := \; I^{+} ( h_{A} ) \, \cap \, I^{-} ( h_{B} ) \\
   W_{3} \;  &  := \; J^{+} (S) \\
   W_{4} \;  &  := \; J^{-} (S)
\end{align}
Let us now suppose to be a statistician living in a Universe
whose Physics sufficentely far from Planck's scale is described
by  a quantum field theory on $ ( M \, , \, g_{a b } ) $,
specified by the set of local observables' algebras $ \{ A_{O}
\}_{O \subseteq M} $ obeying Dimock's  axioms (i.e. Dimock
generalization to curved space-time of Haag-Kastler's axioms)
 \cite{Dimock-80}, \cite{Haag-96}, \cite{Verch-02} whose
world-line is a flow line of the above Killing vector field
$X^{a}$.

Supposing he can access only the state of affairs concerning the
physical observables localized in $ W_{1} $ his objective is to
make a statistical inference concerning the state of affairs
outside $ W_{1} $.

Denoting by:
\begin{equation}
  A_{accessible} \; := \; A_{W_{1}}
\end{equation}
the algebra of observables that is accessible to him, his
objective is to estimate the true state $ \omega \in S( A^{W}_{( M
\, ,\, g_{a b })})  $ of the noncommutative probability space $(
A^{W}_{( M \, ,\, g_{a b })} \, , \omega ) $ describing the
Universe in the assumed classical-background approximation, $
A^{W}_{( M \, ,\, g_{a b })} $ denoting the Weyl algebra of $ ( M
\, , \, g_{a b } ) $, from the knowledge of the information
accessible to him, codified by the \textbf{accessible state}
defined as the restriction of $ \omega $ to the
\textbf{accessible algebra}:
\begin{equation}
  \omega_{accessible} \; := \; \omega \, |_{ A_{accessible} } \;
  \in S ( A_{accessible} )
\end{equation}
Noncommutative Bayesian Statistical Theory, as described  in the
 previous section, would prescribe to him to adopt the following recipe:
\begin{enumerate}
  \item before using even the partial information that is
  accessible to him, the better \textbf{a-priori estimation} of  $ \omega $
  consists in introducing no bias, assuming that:
\begin{equation*}
  \omega_{A \; PRIORI} \; := \; \tau_{unbiased}
\end{equation*}
where $ \tau_{unbiased} $ is the noncommutative unbiased
probability distribution over $ A^{W}_{( M \, ,\, g_{a b })} $,
namely the tracial state on it.
  \item the adoption of the available information may be encoded
  in the passage from the \textbf{a-priori estimation} to the \textbf{a-posteriori
  estimation}  of $ \omega $ specified by the \textbf{noncommutative Bayes
  rule}:
\begin{equation}
   \omega_{A \; PRIORI} ( \cdot ) \, = \, \tau_{unbiased} ( \cdot ) \; \rightarrow \\
   \omega_{A \;
   POSTERIORI} ( \cdot ) \; := \;  \omega_{accessible} ( E_{A \; PRIORI} \cdot )
\end{equation}
where $ E_{A \; PRIORI} \, : \, A^{W}_{( M \, ,\, g_{a b })}
\mapsto A_{accessible} $ is the \textbf{conditional expectation
w.r.t. $ A_{accessible} $ $\omega_{A \; PRIORI}$-invariant}
\end{enumerate}
Let us observe, first of all, that, according to Takesaki Theorem,
the existence of the involved conditional expectation and, hence,
the feasibility of the Bayesian statistical inference, requires
the assumption of the following \textbf{modular constraint}:
\begin{equation} \label{eq:modular constraint}
  \sigma^{\omega_{A \; PRIORI}}_{t} ( a ) \; \in \; A_{accessible}
  \; \;
   \forall a \in A_{accessible} \, , \, \forall t \in {\mathbb{R}}
\end{equation}
 Let us observe, furthermore, that since the Weyl's algebra $ A^{W}_{( M \, ,\, g_{a b })} $ of $ ( M
\, , \, g_{a b } ) $ is generally not finite, the unbiased
noncommutative probability measure $ \tau_{unbiased} $ doesn't
exist.

It must be observed, at this point, that there exists, anyway, an
a priori information that the statistician can adopt: the fact
that the expectation value $  < T_{a b} > $  of the stress-energy
operator $ T_{a b} $ must be well-defined in order of making the
back-reaction's semi-classical Einstein equation:
\begin{equation}
  G_{a b } \; = \; 8 \pi  < T_{a b} >
\end{equation}
well-defined too, resulting in the condition that $ \omega  $ is
an \textbf{Hadamard state} over the Weyl's algebra $ A^{W}_{( M \,
,\, g_{a b })}$ of $ ( M \, , \, g_{a b } ) $:
\begin{equation}
 \omega \; \in \;   S_{H} ( A^{W}_{( M \, ,\, g_{a b })} )
\end{equation}

Consequentially it is natural for the statistician to assume that
$ \omega_{A \; PRIORI} $ is an Hadamard state too.
\begin{equation}
 \omega_{A \; PRIORI} \; \in \;  S_{H} ( A^{W}_{( M \, ,\, g_{a b })} )
\end{equation}

Furthermore he  a-priori knows that some information about the
observables not accessible to him may be recovered by the
information concerning $ A_{accessible} $ through the condition:
\begin{equation}
\exists \{ \alpha_{g} \} \in GR-INN [ Is(M , g_{ab}), A^{W}_{( M
\, ,\, g_{a b })}] \; :
  \alpha_{g} ( A_{O} ) \, = \, A_{ g \, O} \; \; \forall O
  \subset W_{1}
\end{equation}
Consequentially it is natural, for the statistician, to choose
the a-priori state as much $Is[(M,g_{a b})]$-invariant as
possible.

To understand how this two constraints concretely work as to the
determination of $ \omega_{A \; PRIORI} $ it is useful to start
analyzing the simpler cases.
\newpage
\section{The Minkowski case}

Let us start analyzing the simpler particular case in which $ ( M
\, , \, g_{a b } ) $  is the Minkowski space-time:
\begin{align}
  M \; & := \;  {\mathbb{R}}^{4} \\
  g_{a b} \; := \; \eta_{a b} \;  & := \; \eta_{\mu \nu} d x^{\mu } \otimes d x^{\nu } \\
  \eta_{\mu \nu} \; & := \; diag( - 1 , 1 , 1 , 1 )
\end{align}
The isometries-group of Minkowski space-time is the Poincar\'{e}
group $ SO(1,3) \, \circledS \, {\mathbb{R}}^{4} $  generated by
the 10 Killing vector fields:
\begin{equation}
  T_{(i)}^{\mu} \; := \; \delta_{i}^{\mu} \; \; i=0, \cdots , 3
\end{equation}
\begin{equation}
  L_{\mu \nu} \; := \; x_{\mu} \partial_{\nu} \, - \, x_{\nu}
  \partial_{\mu} \; \; \nu \, > \mu  = 0 \, \cdots , 3
\end{equation}
Let us then observe that the surface:
\begin{equation}
  S \; := \; \{ x^{\mu} \in {\mathbb{R}}^{4} \; : \; x^{0} \, =  \, x^{1} \, = \, 0 \}
\end{equation}
is a bifurcate Killing horizon for the Killing vector field:
\begin{equation}
  X^{a} \; := \; L_{0 1}
\end{equation}
generating boosts in the direction $ x^{1} $. Let us denote by
 $ \alpha_{t} $ the inner automorphisms' group representing the
one-dimensional subgroup $ i_{t} $ of $ Is[ ( {\mathbb{R}}^{4} \,
, \, \eta_{a b } ) ] $ generated by $ X^{a} $.

Since the domain of dependence D(S) of the surface S is such that:
\begin{equation}
  D(S) \; = \; {\mathbb{R}}^{4}
\end{equation}
S is a Cauchy surface, so that, according to the general analysis
previously introduced, one has the splitting of the Minkowski
space-time in the four wedges specified by eq.\ref{eq:splitting in
four wedge} with:
\begin{align}
   h_{A} \;  &  \; = \; \{ x^{\mu} \in {\mathbb{R}}^{4} \; : \; x^{0} \, =  \, x^{1}  \} \\
   h_{B} \;  &  \; = \; \{ x^{\mu} \in {\mathbb{R}}^{4} \; : \; x^{0} \, = - \, x^{1}
   \} \\
   W_{1} \;  &  \; = \; \{ x^{\mu} \in {\mathbb{R}}^{4} \; : \;  | x^{1} | \, < \, x^{0} \, , \, x^{0}
   > 0 \} \\
   W_{2} \;  &  \; = \; \{ x^{\mu} \in {\mathbb{R}}^{4} \; : \;  | x^{1} | \, < \, x^{0} \, , \, x^{0} < 0
   \} \\
   W_{3} \;  &  \; = \; \{ x^{\mu} \in {\mathbb{R}}^{4} \; : \;  | x^{1} | \, > \, x^{0} \, , \, x^{1} > 0
   \} \\
   W_{4} \;  &  \; = \; \{ x^{\mu} \in {\mathbb{R}}^{4} \; : \;  | x^{1} | \, > \, x^{0} \, , \, x^{1} < 0
   \}
\end{align}
Since $ X^{a} $ is time-like in the two wedges  $ W_{1} $ and $
W_{2} $ its flow $ i_{t} $  represent possible world-lines of a
massive observer such as our statistician; we will suppose,
precisely, that the statistician's world-line is an integral curve
of $ X^{a} $ contained in $ W_{1} $.

Following the condition enunciated in the last section, among the
possible Hadamard states that our statistician may choose as
a-priori state, the more natural one is the restriction to $
A_{W_{1}} $ of the only $Is( {\mathbb{R}}^{4} , \eta_{a b} )
$-invariant one, i.e. the vacuum state $ \omega_{(0)} $:
\begin{equation}
  \omega_{A PRIORI} \; := \; \omega_{(0)} | _{A_{W_{1}}} \; \in S(
  A_{W_{1}})
\end{equation}
The Unruh effect, consisting in the fact that, in the case $
\omega = \omega_{(0)} $ in which the state to estimate is the
vacuum one, such a vacuum state appears to the statistician
following the flow $ i_{t} $ of $ L_{01} $ as a thermal bath ,
has been explained by Geoffrey Sewell in terms of Modular Theory
through the Bisognano-Wichmann theorem
\cite{Narnhofer-Peter-Thirring-96}, \cite{Haag-96} stating that $
\omega_{A PRIORI} $ is an $\alpha_{t} $- KMS-state at $ \beta = 2
\pi  $.

Let us now recall that the feasibility of the statistical
inferential problem is itself ruled by the modular group of $
\omega_{A PRIORI} $ through the \textbf{modular constraint} of
eq\ref{eq:modular constraint} whose satisfaction, in the present
case:
\begin{equation}
  \sigma_{t}^{\omega_{0}} (a)  \in  A_{W_{1}} \; \; \forall a \in
  A_{W_{1}}
\end{equation}
should follows by the $i_{t}$-invariance of $ \omega_{0} $, by
the fact that $ ( W_{1} , \eta_{a b }|_{\Gamma ( T^{(0,2) W_{1} })
} ) $ is a globally-hyperbolic space-time for its own and by the
fact that:
\begin{equation}
 \alpha_{t} A_{O} \; = \; A_{i_{t}O} \; \; \forall O \subset W_{1}
\end{equation}

Modular Theory tells us, furthermore, that $ \omega_{A \; PRIORI}
$ is a $ \sigma^{\omega_{A \; PRIORI}}_{- \, t} $- KMS state at $
\beta = 1 $.

This double role of the modular group $ \sigma^{\omega_{A \;
PRIORI}}_{t} $ could suggest an interpretation of the Unruh
effect (stating that, in the case $ \omega = \omega_{(0)} $ in
which the state to estimate is the vacuum one, our accelerated
statistician feels a positive temperature) in terms of the
Noncommutative Bayesian Statistical Inference he performs about
the whole noncommutative probability space $ ( A^{W}_{(
{\mathbb{R}}^{4} \, ,\, \eta_{a b })} \, , \, \omega ) $ having
access only to the local information of $  A_{W_{1}} $.

Such a strategy of statistical inference is codified through the
following \textbf{modified Bayes recipe}:
\begin{enumerate}
  \item before making use even of the partial information that is
  accessible to him, the better \textbf{a-priori estimation} of  $ \omega  $
  consists in  assuming as a-priori-state the restriction of the vacuum state
  to the accessible algebra:
\begin{equation*}
  \omega_{A \; PRIORI} \; := \; \omega_{(0)} |_{A_{W_{1}}}
\end{equation*}
  \item the adoption of the available information may be encoded
  in the passage from the \textbf{a-priori estimation} to the \textbf{a-posteriori
  estimation}  of $ \omega $ specified by the \textbf{noncommutative Bayes
  rule}:
\begin{equation}
   \omega_{A \; PRIORI} ( \cdot ) \, = \, \omega_{(0)} |_{A_{W_{1}}} ( \cdot ) \; \rightarrow \\
   \omega_{A \;
   POSTERIORI} ( \cdot ) \; := \;  \omega_{accessible} ( E_{A \; PRIORI} \cdot )
\end{equation}
where $ E_{A \; PRIORI} \, : \, A^{W}_{( {\mathbb{R}}^{4} \, ,\,
\eta_{a b })} \mapsto A_{W_{1}} $ is the \textbf{conditional
expectation w.r.t. $ A_{W_{1}} $ $\omega_{A \; PRIORI}$-invariant}
\end{enumerate}

\newpage
\section{The De Sitter case}
Let us then pass to analyze  the case in which $ ( M \, , \, g_{a
b } ) $ is the De Sitter space-time of unit radius
\cite{Narnhofer-Peter-Thirring-96}:
\begin{align}
  M \; &  \; :=  \; \{ x^{\mu}  \in {\mathbb{R}}^{5} \,  : \, \eta_{\mu \nu} x^{\mu} x^{\nu} \; = \; 1 \} \\
   g_{a  b}  \;  &  := \; i^{\star} \eta_{A B} \\
   \eta_{A B} \; & :=  \eta_{\mu \nu} d x^{\mu} \otimes d x^{\nu} \\
  \eta_{\mu \nu} \; & := diag(-1
   ,1,1,1,1)
\end{align}
namely the hyperboloid of unit radius  embedded in the
(1,4)-Minkowskian space-time $ ( {\mathbb{R}}^{5} \, , \, \eta_{A
B} ) $ endowed with the  lorentzian metric induced by the
inclusion (identity) embedding $ i : M \mapsto {\mathbb{R}}^{5} \;
: \; i(p) := p \; \forall p \in M $

The isometries-group of $ ( {\mathbb{R}}^{5} \, , \, \eta_{A B} )
$:
\begin{equation}
  Is[ ({\mathbb{R}}^{5} \, , \, \eta_{A B} )]| \; = \;  SO(1,4) \circledS
{\mathbb{R}}^{5}
\end{equation}
is generated by the 15 Killing vector fields:
\begin{equation}
  T_{(i)}^{\mu} \; := \; \delta_{i}^{\mu}  \; \; i =
  0 , \cdots, 4
\end{equation}
\begin{equation}
  L_{\mu \nu}   \; := \;  x_{\mu} \partial_{\nu} \, - \, x_{\nu}
  \partial_{\mu} \; \;  \nu  > \mu  = 0 \, \cdots , 4
\end{equation}
While:
\begin{equation}
  i_{\star} L_{\mu \nu}  \in \Gamma (TM) \; \;   \nu  > \mu  = 0 \, \cdots , 4
\end{equation}
one has that:
\begin{equation}
  i_{\star} T_{(i)}^{\mu} \notin \Gamma (TM) \; \; i =
  0 , \cdots, 4
\end{equation}
It follows that the isometry group of the  De Sitter space-time $
( M \, , \, g_{a b} ) $:
\begin{equation}
  Is[ ( M \, , \, g_{a b} )  ] \; = \; SO(1,4)
\end{equation}
is generated by the 10 Killing vector fields:
\begin{equation}
  i_{\star} L_{\mu \nu}  \in \Gamma (TM)  \; \; \nu  > \mu  = 0  , \cdots , 4
\end{equation}

Let us then observe that:
\begin{equation}
  S \; := \; \{ x^{\mu} \in {\mathbb{R}}^{5} \; : \; x^{0} \, =  \, x^{1} \, = \, 0 \}
\end{equation}
is a bifurcate Killing horizon for the  Killing vector field $
L_{01} $ of $( {\mathbb{R}}^{5} \, , \, \eta_{A B} ) $  generating
boosts in the direction $ x^{1} $.

Since the domain of dependence D(S) of the surface S is such that:
\begin{equation}
  D(S) \; = \; {\mathbb{R}}^{5}
\end{equation}
, i.e. S is a Cauchy surface of $( {\mathbb{R}}^{5} \, , \,
\eta_{A B} )$, according to the general analysis previously
introduced one has the splitting of the (1,4)-Minkowski
space-time in four wedges specified by eq.\ref{eq:splitting in
four wedge} with:
\begin{align}
   h_{A} \;  &  \; = \; \{ x^{\mu} \in {\mathbb{R}}^{5} \; : \; x^{0} \, =  \, x^{1}  \} \\
   h_{B} \;  &  \; = \; \{ x^{\mu} \in {\mathbb{R}}^{5} \; : \; x^{0} \, = - \,
   x^{1} \} \\
   W_{1} \;  &  \; = \; \{ x^{\mu} \in {\mathbb{R}}^{5} \; : \;  | x^{1} | \, < \, x^{0} \, , \, x^{0}
   > 0 \} \\
   W_{2} \;  &  \; = \; \{ x^{\mu} \in {\mathbb{R}}^{5} \; : \;  | x^{1} | \, < \, x^{0} \, , \, x^{0} < 0
   \} \\
   W_{3} \;  &  \; = \; \{ x^{\mu} \in {\mathbb{R}}^{5} \; : \;  | x^{1} | \, > \, x^{0} \, , \, x^{1} > 0
   \} \\
   W_{4} \;  &  \; = \; \{ x^{\mu} \in {\mathbb{R}}^{5} \; : \;  | x^{1} | \, > \, x^{0} \, , \, x^{1} < 0
   \}
\end{align}
It follows that $S |_{M} $ is a bifurcate Killing horizon for the
Killing vector field:
\begin{equation}
  X^{a} \; := \; i_{\star} L_{01}
\end{equation}
of  $( M \, , \, g_{ab})$. Since $ S|_{M} $ is a Cauchy surface
of $( M \, , \, g_{ab} )$ :
\begin{equation}
  D( S|_{M} ) \; = \; M
\end{equation}
one has the splitting of $( M \, , \, g_{ab} )$ into the four
wedges specified by eq.\ref{eq:splitting in four wedge} with:
\begin{equation}
  H_{i} \; := \; W_{i} |_{M} \; \; i = 1 , \cdots , 4
\end{equation}
Since the $ Is(M , g_{a b})$-invariance selects again a single
state, the vacuum state $ \omega_{0} $, among the Hadamard ones,
the prescribed conditions for the selection of the a-priori state
lead us to the \textbf{modified Bayes recipe}:
\begin{enumerate}
  \item before making use even of the partial information that is
  accessible to him, the better \textbf{a-priori estimation} of  $ \omega  $
  consists in  assuming as a-priori-state the restriction of the vacuum state
  to the accessible algebra:
\begin{equation*}
  \omega_{A \; PRIORI} \; := \; \omega_{(0)} |_{A_{H_{1}}}
\end{equation*}
  \item the adoption of the available information may be encoded
  in the passage from the \textbf{a-priori estimation} to the \textbf{a-posteriori
  estimation}  of $ \omega $ specified by the \textbf{noncommutative Bayes
  rule}:
\begin{equation}
   \omega_{A \; PRIORI} ( \cdot ) \, = \, \omega_{(0)} |_{A_{H_{1}}} ( \cdot ) \; \rightarrow \\
   \omega_{A \;
   POSTERIORI} ( \cdot ) \; := \;  \omega_{accessible} ( E_{A \; PRIORI} \cdot )
\end{equation}
where $ E_{A \; PRIORI} \, : \, A^{W}_{( M \, ,\, g_{a b })}
\mapsto A_{W_{1}} $ is the \textbf{conditional expectation w.r.t.
$ A_{W_{1}} $ $\omega_{A \; PRIORI}$-invariant}
\end{enumerate}

\medskip

Let us denote by $ \alpha_{t} $ the inner automorphisms' group
representing the one-dimensional subgroup $ i_{t} $  of $ Is[ ( M
\, , \, g_{a b } ) ] $ generated by $ X^{a} $.

The Unruh effect, consisting in the fact that, in the case $
\omega = \omega_{(0)} $ in which the state to estimate is the
vacuum one, such a vacuum state appears to the statistician
following the flow  of $ X^{a} $ as a thermal bath, has been
recasted by Figari, H \"{o}egh-Krohn an Nappi and later by  Bros
and Moschella  in terms of Modular Theory through the
Bisognano-Wichmann theorem \cite{Narnhofer-Peter-Thirring-96},
\cite{Haag-96} stating that $ \omega_{A PRIORI} $ is an
$\alpha_{t} $- KMS-state at $ \beta = 2 \pi  $.

Let us now recall that, exactly as in the Minkowskian case, the
feasibility of the statistical inferential problem is itself
ruled by the modular group of $ \omega_{A PRIORI} $ through the
\textbf{modular constraint} of eq\ref{eq:modular constraint}
whose satisfaction, in the present case:
\begin{equation}
  \sigma_{t}^{\omega_{0}} (a)  \in  A_{H_{1}} \; \; \forall a \in
  A_{H_{1}}
\end{equation}
should follows by the $i_{t}$-invariance of $ \omega_{0} $, by
the fact that $ ( H_{1} , g_{a b }|_{\Gamma ( T^{(0,2) H_{1} }) }
) $ is a globally-hyperbolic space-time for its own, and by the
fact that:
\begin{equation}
 \alpha_{t} A_{O} \; = \; A_{i_{t}O} \; \; \forall O \subset H_{1}
\end{equation}

Modular Theory, furthermore, tells us that  $ \omega_{A \;
PRIORI} $ is a $ \sigma^{\omega_{A \; PRIORI}}_{- \, t} $- KMS
state at $ \beta = 1 $.

So, once again, this double role of the modular group $
\sigma^{\omega_{A \; PRIORI}}_{t} $ could suggest an
interpretation of the Unruh effect in terms of the Noncommutative
Bayesian Statistical Inference he performs about the whole
noncommutative probability space $ ( A^{W}_{( M \, ,\, g_{a b })}
\, , \, \omega ) $ having access only to the local information of
$ A_{H_{1}} $.

\end{document}